\documentclass[aps,twocolumn,showpacs,preprintnumbers,nofootinbib,prl,superscriptaddress,10pt]{revtex4-2}

\makeatletter
\def\l@subsubsection#1#2{}
\def\l@subsubsubsection#1#2{}
\makeatother

\usepackage{graphicx,amssymb,amsmath,amsthm,amsfonts,epsfig,epsf,fixmath}
\usepackage[dvipsnames]{xcolor}
\usepackage{epstopdf}

\usepackage{aas_macros}
\usepackage{bm}
\usepackage{comment}
\usepackage{dcolumn}
\usepackage{latexsym}
\usepackage{rotating}
\usepackage{longtable}

\usepackage{enumerate}
\usepackage{tensor,multirow}
\usepackage{url}
\usepackage[linktocpage]{hyperref}
\usepackage{soul}
\usepackage{physics}

\usepackage{ulem} 

\newcommand{\tn}{\textnormal}

\newcommand{\GSSI}{Gran Sasso Science Institute (GSSI), I-67100 L’Aquila, Italy}
\newcommand{\GranSasso}{INFN, Laboratori Nazionali del Gran Sasso, I-67100 Assergi, Italy}

\begin{document}
\title{
Probing fundamental physics with Extreme Mass Ratio Inspirals:\\
full Bayesian inference for scalar charge
}

\author{Lorenzo Speri}
\affiliation{Max Planck Institute for Gravitational Physics (Albert Einstein Institute), Am M{\"u}hlenberg 1, Potsdam 14476, Germany}
\author{Susanna Barsanti}
\affiliation{School of Mathematical Sciences \& School of Physics and Astronomy, University of Nottingham, University Park, Nottingham, NG7 2RD, UK}
\affiliation{Nottingham Centre of Gravity, University of Nottingham, University Park, Nottingham, NG7 2RD, UK}
\author{Andrea Maselli}
\affiliation{\GSSI}
\affiliation{\GranSasso}
\author{Thomas P. Sotiriou}
\affiliation{School of Mathematical Sciences \& School of Physics and Astronomy, University of Nottingham, University Park, Nottingham, NG7 2RD, UK}
\affiliation{Nottingham Centre of Gravity, University of Nottingham, University Park, Nottingham, NG7 2RD, UK}
\author{Niels Warburton}
\affiliation{School of Mathematics and Statistics, University College Dublin, Belfield, Dublin 4, Ireland}
\author{Maarten van de Meent}
\affiliation{Niels Bohr International Academy, Niels Bohr Institute, Blegdamsvej 17, 2100 Copenhagen, Denmark}
\affiliation{Max Planck Institute for Gravitational Physics (Albert Einstein Institute), Am M{\"u}hlenberg 1, Potsdam 14476, Germany}
\author{Alvin J. K. Chua}
\affiliation{Department of Physics, National University of Singapore, Singapore 117551}
\affiliation{Department of Mathematics, National University of Singapore, Singapore 119076}
\author{Ollie Burke}
\affiliation{Laboratoire des 2 Infinis -- Toulouse (L2IT-IN2P3), Universitè de Toulouse, CNRS, UPS, F-31062 Toulouse Cedex 9, France}
\author{Jonathan Gair}
\affiliation{Max Planck Institute for Gravitational Physics (Albert Einstein Institute), Am M{\"u}hlenberg 1, Potsdam 14476, Germany}

\begin{abstract}
Extreme Mass Ratio Inspirals (EMRIs) are key sources for the future space-based gravitational wave detector LISA, and are considered promising probes of fundamental physics. Here, we present the first complete Bayesian analysis of EMRI signals in theories with an additional massless scalar, which could arise in an extension of General Relativity or of the Standard Model of Particle Physics. We develop a waveform model accurate at adiabatic order for equatorial eccentric orbits around spinning black holes. Using full Bayesian inference, we forecast LISA's ability to probe the presence of new fundamental fields with EMRI observations.
\end{abstract}


\maketitle

\noindent{{\bf{Introduction.}}} 
In the last decade, gravitational wave (GW) astronomy has revolutionized our ability to observe the Universe, offering unique opportunities to test the nature of gravity and search for new fundamental fields in previously unexplored regimes \cite{Barack:2018yly,Barausse:2020rsu,LISA:2022kgy}. Observations with gravitational-wave detectors allow us to probe the highly dynamic and strong-field regime of compact binary coalescences.

Ground-based interferometers have paved the way for searches for new physics beyond General Relativity (GR) or the Standard Model (SM) \cite{LIGOScientific:2021sio}. The constraining power of current facilities is limited to signal-to-noise ratio up to $30$ \cite{GW170817}, comparable mass binaries, with the most asymmetric system detected so far having a mass ratio of {\( q = \mu/M\sim 1/10 \)}, with GW190814 \cite{LIGOScientific:2020stg,GW230529,LIGOScientific:2020zkf}, {where we defined $M, \mu$ the primary and secondary component masses, respectively}.
The next generation of ground-based detectors, such as the Einstein Telescope \cite{Punturo:2010zz} and Cosmic Explorer \cite{Reitze:2019iox,2015PhRvD..91h2001D}, along with space-based missions like the Laser Interferometer Space Antenna (LISA) \cite{Colpi:2024xhw}, TianQin \cite{TianQin:2020hid}, Taiji \cite{Taiji}, and Lunar Gravitational Wave Antenna \cite{Ajith:2024mie}, are expected to observe sources across a broader mass range. Probing fundamental physics is featured prominently in  science cases for future detectors~\cite{Sathyaprakash:2019yqt,Kalogera:2021bya,Barausse:2020rsu,LISA:2022kgy,Colpi:2024xhw}.

LISA is expected to detect gravitational waves from Extreme Mass Ratio Inspirals (EMRIs):  binary systems composed of stellar-mass compact objects (the secondary), inspiralling into massive black holes at the center of galaxies (the primary). 
Due to their small mass ratios $q\lesssim 10^{-4}$, 
EMRIs perform tens of thousands of cycles on highly relativistic trajectories with large inclination and orbital eccentricities within the LISA sensitivity band.
The orbit complexity results in gravitational signals with multiple harmonics.
The rich harmonic content and the many cycles of EMRI signals allow sub-percent parameter measurement precision, rendering EMRIs natural laboratories to 
test gravity and offering a unique opportunity to probe fundamental physics in unprecedented regimes \cite{Barausse:2020rsu}. These same characteristics also make the modeling and generation of EMRI waveforms particularly challenging.

A variety of studies have investigated the scientific potential of such sources to test the nature of black holes (BHs) \cite{Barack:2006pq,Sopuerta:2009iy,Babak:2017tow,Fransen:2022jtw,Raposo:2018xkf,Bena:2020see,Bianchi:2020bxa,Loutrel:2022ant,Piovano:2020ooe,Pani:2019cyc,Piovano:2022ojl,Datta:2019epe,Datta:2019euh,Maggio:2021uge,Pani:2010em,Macedo:2013qea,Destounis:2023khj,Datta:2019epe,Datta:2019euh,Liu:2020ghq,Burke:2020vvk,Rahman:2021eay,Maggio:2021uge,Miller:2024rca,Zi:2024dpi,Cardenas-Avendano:2024mqp,Zi:2023qfk,Chua:2018yng,Gair:2011ym}, the propagation speed of gravity~\cite{Liu:2023onj,Saltas:2023qec}, and the existence of new fundamental fields \cite{Cardoso:2011xi,Yunes:2011aa,Pani:2011xj,Canizares:2012is,Hannuksela:2018izj,Hannuksela:2019vip,Maselli:2020zgv,Maselli:2021men,Barsanti:2022ana,Barsanti:2022vvl,DellaRocca:2024pnm,Liang:2022gdk,Zhang:2023vok,Zi:2022hcc,Lestingi:2023ovn,Collodel:2021jwi, Duque:2023cac,Brito:2023pyl,Chen:2024ery,Fell:2023mtf}.
While testing gravity with EMRIs is a key science goal for LISA, calculations of waveform models beyond GR are in their infancy. The majority of works carried out so far have 
resorted to ad-hoc modifications of GR templates, adopting hybrid schemes based, for 
example, on post-Newtonian (PN) expansions \cite{Speri:2022upm}. Moreover, such studies have mainly focused 
on the potential of EMRI observations to identify deviations from the spacetime of the 
primary, forecasting --- in some cases --- constraints on such parameters from 
LISA observations \cite{Barack:2006pq,Canizares:2012is}.

Contrary to the standard lore, which has the EMRI acting as a probe for the spacetime of the primary, Ref.~\cite{Maselli:2020zgv} demonstrated that changes from the Kerr metric of the primary can be neglected at leading order in the mass ratio for a large family of theories with scalar fields non-minimally coupled to gravity. It is instead the \textit{scalar charge} of the much lighter secondary that can leave a significant imprint on the GW emission. 
This is because for massless scalars, the scalar charge, if any, is inversely proportional to the square of the mass of the black hole \cite{Sotiriou:2013qea,Sotiriou:2014pfa,Thaalba:2022bnt,Saravani:2019xwx,Spiers:2023cva}.
The framework developed in Ref.~\cite{Maselli:2020zgv} allows the construction of waveforms that are correct at the leading adiabatic order (first order in the mass ratio) and for which deviations from GR are uniquely determined by the scalar charge of the secondary. This framework has recently been framed into a consistent approach to compute post-adiabatic (second order in the mass ratio) waveform corrections to the GR baseline model \cite{Spiers:2023cva}. 

This formalism has been exploited to study changes in the GW fluxes for binaries on eccentric equatorial \cite{Barsanti:2022ana} and circular inclined orbits \cite{DellaRocca:2024pnm,Warburton:2014bya}  for massless scalar fields, and for circular equatorial inspirals in the case of massive scalars \cite{Barsanti:2022vvl}. Preliminary analyses have also assessed LISA's capability to infer the measurement precision of the scalar charge using Fisher information matrix calculations \cite{Maselli:2021men,Barsanti:2022vvl,Tan:2024utr,Zhang:2022rfr,Guo:2022euk}.

Here, we provide the first implementation of adiabatic waveforms for EMRIs with scalar fields, in eccentric equatorial orbits around spinning BHs.
{While EMRI formation scenarios are generally expected to be generic, we consider this implementation an important step forward in modeling EMRIs within beyond-GR frameworks.}
We perform a Bayesian analysis on all the waveform parameters to {forecast the power of single EMRI observations} to detect the scalar charge, and, therefore, probe deviations from General Relativity and the Standard Model. {The waveform model and the Bayesian analysis codes are available at \url{https://github.com/lorenzsp/testGRwEMRIs}}

Our results show that a single EMRI is able to constrain the scalar charge of the secondary, with precision of 10\% in a theory-agnostic 
way, i.e., independent of the origin of the scalar field.
Moreover, if one selects a specific theory, a constraint of the charge can be converted into a bound on a coupling constant of this theory, which controls deviations from GR. As a characteristic example, we consider here the case of linear-Gauss--Bonnet gravity.

\noindent{{\bf{EMRIs and fundamental fields.}}} 
In this Section, we briefly summarise the theoretical approach we use to model EMRIs with scalar fields \cite{Maselli:2020zgv,Maselli:2021men,Barsanti:2022ana, DellaRocca:2024pnm}. We refer the reader to 
\cite{Spiers:2023cva} for a detailed description of the formalism and its extension within a Self-Force 
(SF) scheme. 

We consider theories with a single massless scalar field $\phi$, non-minimally coupled to the metric 
tensor ${\bf g}$, described by the following action
\begin{equation}
   \int \dd^4 x \frac{\sqrt{-g}}{16 \pi} \left[R - \frac{1}{2} \partial_\mu \phi \partial^{\mu} \phi \right] 
   + \alpha_c S_{\rm c} \left[\textbf{g}, \phi\right] + S_{\rm m}\left[\textbf{g}, \phi, \Psi\right]\ ,
    \label{eq:action}
\end{equation}
where $R$ is the Ricci scalar and $g$ is the metric determinant. $S_{\rm m}$ describes the dynamics of the matter fields $\Psi$. The action $\alpha_c S_{\rm c}$ describes (each of) the scalar field interactions, with the constant $\alpha_c$ 
having dimensions $({\rm mass})^n$, with $n>1 $. In physical units, this corresponds to interactions that 
are suppressed by a characteristic energy scale \cite{Saravani:2019xwx}. 
Varying with respect to the metric and the scalar field, we obtain the equations for the fields
\begin{equation}
    G_{\mu\nu} =   8 \pi T^{\rm scal}_{\mu \nu}+\alpha_c T^{\rm c}_{\mu\nu}+ T^{\rm m}_{\mu\nu}\quad\ ,\quad \Box \phi  = T^{\rm c}+ T^{\rm m} \ ,\label{eq:fieldseq}
\end{equation}
where $\square=\nabla_\mu\nabla^\mu$, 
$T^{\rm scal}_{\mu \nu} = \frac{1}{16 \pi } \left[\partial_{\mu} \phi\partial_{\nu} \phi - \frac{1}{2}g_{\mu\nu}(\partial \phi)^2\right]$ 
and 
\begin{equation}
T^{\rm c,m}_{\mu\nu} =- \frac{16 \pi}{\sqrt{-g}}\frac{\delta S_{\rm c,m}}{\delta g^{\mu \nu}}   \quad\ ,\quad
T^{\rm c,m} =- \frac{16 \pi}{\sqrt{-g}}\frac{\delta S_{\rm c,m}}{\delta \phi}\ . 
\end{equation}
Since we are considering a massless scalar, we will assume that $S_c$ respects shift symmetry, $\phi \to \phi+$ constant. However, our approach can be generalised to light scalars \cite{Barsanti:2022vvl}. 
We focus on  EMRIs in which the primary is a BH of mass $M$, and assume that solutions in theories controlled by \eqref{eq:action} are continuously connected to GR solutions for $\alpha_c\rightarrow 0$. For shift-symmetric scalars, the scalar charge for black holes, if any \cite{1970CMaPh..19..276C,PhysRevD.51.R6608,Hawking72,Capuano:2023yyh,Sotiriou:2011dz,Hui:2012qt}, is fixed in terms of their mass, spin and the coupling constants of the theory \cite{Saravani:2019xwx}. Hence, {the primary mass} $M$ and {the coupling constant} $\alpha_c$ are the only meaningful physical scales of this problem. Their ratio can be expressed as
\begin{equation}
\zeta=\frac{\alpha_c}{M^n}=q^n\frac{\alpha_c}{\mu^n}\ .\label{eq:zetapar}
\end{equation}
Existing bounds already require that  $\alpha_c/\mu^n\lesssim \mathcal{O}(1)$ \cite{Nair:2019iur}, so $\zeta$ is order $q^n$. One can then use $q$ 
as a single bookkeeping parameter, as for the SF approach in GR. 
  
This introduces several simplifications in the description of EMRIs~\footnote{The approach extends to less asymmetric binaries, like Intermediate Mass Ratio 
Inspirals, so long as $\zeta$ remains a perturbative parameter.}. 
 Indeed, by expanding the metric and the scalar field in powers of $q$, 
\begin{align}
g_{\mu\nu}=g_{\mu\nu}^{(0)}+q h^{(1)}_{\mu\nu}+\ldots\ \ ,  
\ \phi=\phi^{(0)}+q \phi^{(1)}+\ldots\ ,   
\end{align} 
it was recently shown how to derive a consistent SF formalism that includes post-adiabatic corrections to the binary dynamics \cite{Spiers:2023cva}. In this paper, we focus on the leading EMRI dissipative contribution, which is fully determined by the linear order perturbations $h_{\mu\nu}^{(1)}$ and $\phi^{(1)}$.
Equation~\eqref{eq:zetapar} implies that: (i) the background spacetime is suitably described by the Kerr metric, with beyond GR deviations being $\mathcal{O}(q^{2n})$ corrections to $g_{\mu\nu}^{(0)}$, (ii) $\phi^{(0)}$ is constant due to the no-hair theorem \cite{1970CMaPh..19..276C,PhysRevD.51.R6608,Hawking72,Capuano:2023yyh,Sotiriou:2011dz,Hui:2012qt} and can be set to zero by a shift, (iii) at  adiabatic order, metric and scalar field perturbations induced by the secondary decouple, leading to a separate set of equations:
\begin{align}
    G^{\alpha \beta}[h^{(1)}_{\alpha\beta}] &= 8 \pi \mu  \int \frac{\delta^{(4)}\left(x-y_p(\lambda)\right)}{\sqrt{-g}}\frac{\dd y^{\alpha}_p}{\dd\lambda} \frac{\dd y^{\beta}_p}{\dd\lambda} \dd\lambda \ ,\label{eq:fieldeqg}\\
    \Box \phi^{(1)} &= - 4 \pi d \,  \mu \int \frac{\delta^{(4)} \left(x - y_p(\lambda)\right) }{\sqrt{-g}}\dd\lambda\ , 
    \label{eq:fieldeqphi}
\end{align}
where $G_{\alpha\beta}$ is the Einstein tensor and $dy_p^\mu/d\lambda$ is the four velocity of the secondary, along its worldline.

%


Eqs.~\eqref{eq:fieldeqg} are identical to GR, while Eq.~\eqref{eq:fieldeqphi} determines the scalar field evolution and depends on the scalar charge of the secondary, $d$, which enters as the only extra EMRI parameter. The solution for $h_{\mu\nu}^{(1)}$ and $\phi^{(1)}$ allows the energy and angular momentum fluxes to be computed for the gravitational (grav) sector, $(\dot{E}_{\rm grav}$, $\dot{L}_{\rm grav})$, and the scalar sector (scal) $(\dot{E}_{\rm scal}$, $\dot{L}_{\rm scal})$. Beyond GR modifications to the EMRI evolution are \textit{uniquely} controlled by the latter. At this order in mass ratio, the gravitational waveform amplitudes are the same as in GR, whereas the waveform phase is affected by the extra channel of emission. 
In fact, if the charge is not zero, the secondary plunges faster than in GR. In this work, we consider systems on equatorial eccentric orbits \cite{Barsanti:2022ana}, such that $\dot{E},\dot{L}$  depend on the semi-latus rectum $p$, on the eccentricity $e$, and on the primary {dimensionless} spin $a$.

For a given theory, there is a mapping between $d$ and the coupling constant $\alpha_c$. This allows constraints on $d$ to be translated into bounds on such coupling. As an example, we will consider here 
linear Gauss Bonnet Gravity (GB) \cite{Sotiriou:2013qea}, for which 
\begin{equation}
\alpha_c S_c=\frac{\alpha}{4}\int d^4 x\frac{\sqrt{-g}}{16\pi}\phi {\cal G}\ ,\label{eq:actionGB}
\end{equation}
where ${\cal G}=R^2-4R_{\mu\nu}R^{\mu\nu}+R_{\alpha\beta\mu\nu}R^{\alpha\beta\mu\nu}$ is the Gauss Bonnet invariant, $R_{\alpha\beta\mu\nu}, R_{\alpha\beta}$ are the Riemann and Ricci tensors, and $\alpha$ has dimensions of mass squared ($n=2$). For this theory  $\alpha\simeq 2d\mu^2$ {in geometrized units. 
Throughout this manuscript, we use geometrized units; however, following standard practice in the literature, we also express constraints on $\sqrt{\alpha}$ in kilometres by restoring the gravitational constant $G_N$ and light speed $c$, yielding $\sqrt{\alpha}[{\rm km}]= \sqrt{2d}\mu G_N/c^2$}\cite{Julie:2019sab}. 
In Appendix~\ref{app:normalization_action}, we {discuss different normalizations of the action present in the literature} and obtain the relation between the scalar charge $d$ and the coupling constant $\sqrt{\alpha}$.

\noindent{{\bf{Waveform modelling and data analysis setup.}}} 
We develop a waveform model for EMRIs \cite{Barack:2018yvs,Pound:2019lzj, Warburton:2021kwk, Wardell:2021fyy,Green:2019nam,Dolan:2021ijg,Upton:2021oxf,Toomani:2021jlo,Osburn:2022bby,Spiers:2023cip,Nasipak:2021qfu,Piovano:2020zin,Mathews:2021rod,Drummond:2022xej,Upton:2023tcv,Drummond:2023loz,Upton:2023tcv} that accounts for the scalar emission and implement it within the \texttt{FastEMRIWaveform} (\texttt{FEW}) package {\cite{Chua:2020stf,Katz:2021yft,Speri:2023jte,Chapman-Bird:2025xtd}}, which allows for fast generation of EMRI templates on Graphics Processing Units (GPUs). 

In the first order in the mass ratio, the orbital evolution is obtained by solving the following system of ordinary differential equations
{\cite{Chapman-Bird:2025xtd}:
\begin{equation}
    \frac{dJ}{dt}  = \frac{q}{M} f_J\quad \ ,\quad  
    \frac{d\Phi_{i}}{d t} = \frac{\Omega_{i}}{M}
    \qquad {J} = \{p, e\} \ ,\label{eq:EMRIODE}
\end{equation}
where the time $t$ scales with $M$ and } $\Omega_{i=r,\phi}$ are the dimensionless fundamental frequencies of the Kerr spacetime, {and $f_{p,e}$ are the orbital-element fluxes} that depend on the BH {dimensionless} spin $a$ and on the {dimensionless} orbital elements $\{p,e\}$ \citep{Fujita:2009bp,Schmidt:2002qk}. 
Equations~\eqref{eq:EMRIODE} can be integrated given the initial conditions for the semi-latus rectum $p_0$, for the eccentricity $e_0$, and for the phases $(\Phi_{\phi 0}$, $\Phi_{r 0})$ (see Appendix~\ref{app:ode_error} for a study of the ordinary differential equations' accuracy). The ordinary differential equation is integrated until we reach the separatrix plus a threshold of 
$0.1M$. We call these plunging orbits.

The orbital-element fluxes, $f_{p,e}$, are written in terms of energy and angular momentum fluxes. For example, for the semi-latus rectum, we have
{
\begin{align}
    f_J &= -\frac{M}{q}\qty[\frac{\partial J}{\partial E} \dot{E} + \frac{\partial J}{\partial L} \dot{L}] \ , \quad J=p,e \ ,\\
    \dot{E} &= \dot{E}_{\rm grav}+ d^2\, \dot{E}_{\rm scal},\quad \, \quad \dot{L}=\dot{L}_{\rm grav}+ d^2 \, \dot{L}_{\rm scal} \, ,
\end{align}
}
where both the gravitational and scalar fluxes are the sum of the horizon and infinity fluxes and have been computed\footnote{Note that the scalar fluxes $\dot{E}_{\rm scal}$ and $\dot{L}_{\rm scal}$ derived from the Black Hole Perturbation Toolkit and in \cite{Warburton:2010eq} must be divided by a factor 4, to account for a different normalization of the scalar field used in the action \eqref{eq:action}.} 
{using the code of \cite{EMRIscalarsrep, Warburton:2010eq} and packages of the Black Hole Perturbation Toolkit \cite{BHPToolkit}, using the mode summation implemented in \cite{Skoupy:2022adh}}.
Angular momentum and energy fluxes are computed on a 3d grid in $(a,p,e)$, and interpolated using Chebyshev polynomials \cite{Lynch:2023gpu} (see Appendix~\ref{app:interpolation} for details).

Once the trajectory is implemented, we can pass it to the Augmented Analytical Kludge (AAK) \citep{Chua:2017ujo} {waveform amplitude model} implemented\footnote{In this study we did not use the relativistic version of FEW since that version of the model is still under revision.} in FEW \citep{Chua:2018woh,Chua:2020stf,michael_l_katz_2020_4005001,Katz:2021yft}. 
The AAK waveform is GPU-accelerated and allows the exploitation of the long-wavelength approximation of the LISA response, providing a generic model for investigating tests of GR. The typical waveform generation speed of this new time-domain model is of order $0.1$ seconds.

Our trajectory is fully relativistic at adiabatic order, and therefore the measurement precision of the 
intrinsic parameters $\Theta_i=(\ln M,\ln \mu, a, p_0, e_0, \Phi_{\phi 0 }, \Phi_{r 0 }, d)$ {is not strongly} 
affected by the choice of the AAK template \cite{Katz:2021yft}. 
{The inclusion of post-adiabatic corrections and their impact on the scalar charge detectability will be explored in a follow-up work, 
exploiting recent advancements in parameter estimation for post-adiabatic waveforms \cite{Warburton:2021kwk,Burke:2023lno}.}

The AAK amplitudes and the LISA response may, however, affect the accurate  reconstruction of the extrinsic parameters $\Theta_e$ defined here by the luminosity distance $d_L$, the polar and azimuthal sky location angles, $(\theta_S,\phi_S)$, and the polar and azimuthal orientation angles $(\theta_K,\phi_K)$ that determine the orientation of the primary spin (both set of angles are expressed with respect to the Solar System barycenter reference frame \cite{Katz:2021yft}). 
We have checked using Fisher matrices and MCMC that our results are unchanged when using the full LISA response, and an early implementation of the relativistic amplitudes. 

%
\begin{figure}
\centering
\includegraphics[width=\columnwidth]{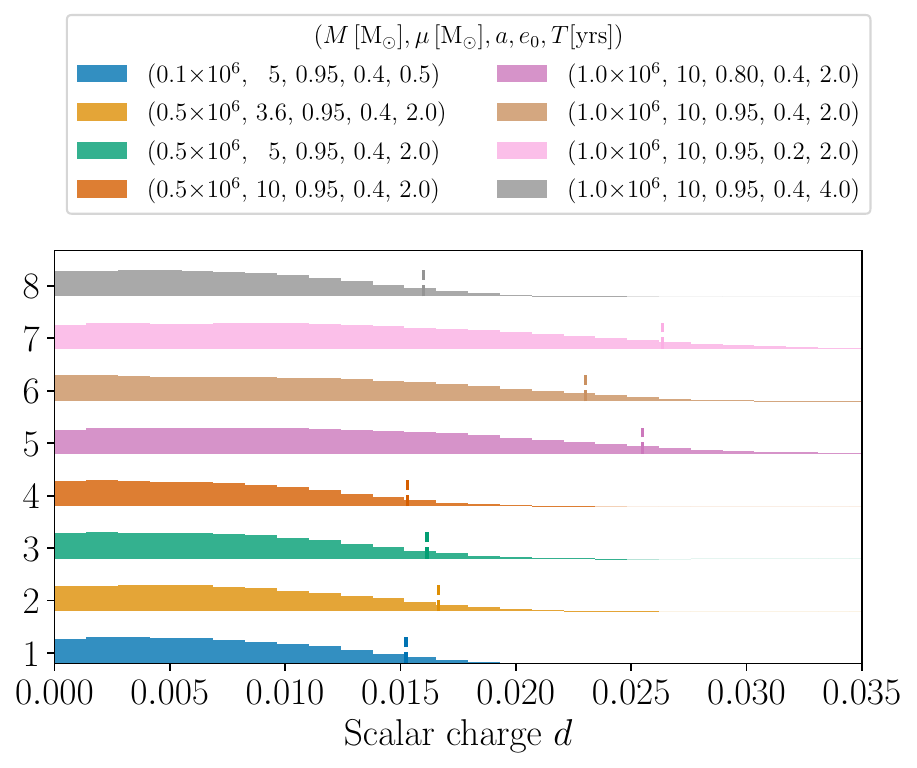}
        \caption{Histograms  of posterior samples of the scalar charge inferred by LISA observations of EMRIs with different orbital configurations of central black hole mass $M$ and {dimensionless} spin $a$, compact object mass $\mu$, initial eccentricity $e_0$ and time to 
        plunge $T$. The colored vertical dashed lines show the one-sided $95\%$ credible interval of the distribution. All EMRI systems are characterized by an SNR of 50.}\label{fig:charge_bound}
    \end{figure}

To forecast the constraints on the scalar charge with EMRI observations we sample over all the intrinsic $\Theta_i$ and extrinsic parameters $\Theta_e$ of a Kerr equatorial eccentric EMRI signal with scalar charge. 
We obtain the 13-dimensional posterior distribution using the package \texttt{Eryn} \cite{Karnesis:2023ras}, which provides a Bayesian inference tool based on Markov Chain Monte Carlo sampling. The technical details about the likelihood, priors, and sampling techniques are extensively discussed in 
Appendix~\ref{app:datanalysis}.
This is the first appearance of a complete Bayesian parameter inference of EMRI systems in these types of orbits.

\noindent{{\bf{Results.}} 
We consider eight different orbital configurations, 
specified by their component masses, the primary {dimensionless} spin, 
the initial eccentricity and semi-latus rectum. We 
assume binaries evolve in the LISA band, plunging over 
a period $T$. This fixes the initial semi-latus rectum 
$p_0$. For each system, the luminosity distance is 
fixed to give SNR$=50$. The number of  EMRIs detectable by LISA 
ranges from a few to hundreds per year \cite{Babak:2017tow}, including 
tens of events with SNRs between 100 and 1000.


We first focus on \textit{agnostic} forecasts of new fundamental 
fields, assessing LISA's ability to constrain the scalar 
charge. To this aim, we study the case in which the 
injected signal is modelled in GR, i.e., assuming $d=0$, 
while the recovery template includes the scalar charge.
This setup allows us to investigate the upper 
bound (or constraint) on $d$, which we define as given by the 
upper 95\% credible interval of the corresponding 
marginalized posterior.

Fig.~\ref{fig:charge_bound} 
shows histograms of the  marginalized posteriors of 
the scalar charge 
for the EMRIs we considered (see Fig.~\ref{fig:full_posterior} of Appendix~\ref{app:datanalysis} for the full posterior). 

Bounds on  $d$ are tighter for large eccentricity 
and primary {dimensionless} spin.
For fixed component masses and evolution time, 
doubling the eccentricity from $e_0=0.2$ to 
$e_0=0.4$ yields a $10\%$ stronger bound on $d$ 
(cf. systems 6 and 7 in Fig.~\ref{fig:charge_bound}).
We find the same level of improvement 
when increasing the BH {dimensionless} spin 
from $a=0.8$ to $a=0.95$ (systems 5 and 6). 
However, the precision might vary differently for 
lower eccentricity binaries, see for instance 
Fig.~9 of \cite{Zhang:2022rfr}.

We also find that increasing the mass ratio 
provides narrower posteriors. Assuming 
$\mu=10\, M_\odot$, if we reduce the primary 
mass by a factor of two, we obtain a bound 
on $d$ that is $\sim50\%$ tighter 
(systems $4$ and $6$). This is primarily 
because a less asymmetric system plunges faster. 
For a fixed evolution time, such 
binaries have larger initial orbital separations, 
where the effect of scalar emission is 
stronger. 
To illustrate this point, we analyze a system with component masses 
$M=10^{5}\,M_\odot$ and $\mu=5\,M_\odot$ (system 1). This system has the largest initial semi-latus rectum $p_0\approx 16 $ and, although we consider only the last half a year before the plunge for computational reasons, it yields the best 95\% upper bound on $d$, $d_{95\%} = 0.015$. 
Fixing the intrinsic source parameters, the 
measurement precision improves for EMRIs 
evolving over a longer timescale. 
For a $10^6\,M_\odot+10\,M_\odot$ system, 
doubling of $T$ improves the constraint 
on $d$ by $60\%$, and increases the number of cycles from $1.1\times 10^5$ to $1.7\times 10^5$, and the semi-latus rectum from $p_0=8.34 $ to $p_0=10 $.


We now explore the case in which \textit{both} 
the injected and the recovery waveforms have a 
non-vanishing scalar charge. We inject a signal 
with $d =0.025$, consistent with the upper bound from GW230529 ($d\approx 0.035$), and study constraints on 
the charge for a $10^5 M_\odot+ 5M_\odot$ EMRI, 
with the same orbital parameters as system 1 
in Fig.~\ref{fig:charge_bound}. 
This binary provides a measurement of the 
charge accurate to $\sim 10\%$, with median 
and 95\% credible intervals of 
$d=0.0244^{+0.006} _{-0.007}$. The marginalized 
posterior of $d$ for this system is shown in 
Fig.~\ref{fig:charge_histogram}.

\begin{figure}[hbpt!]
    \centering \includegraphics[width=0.9\columnwidth]{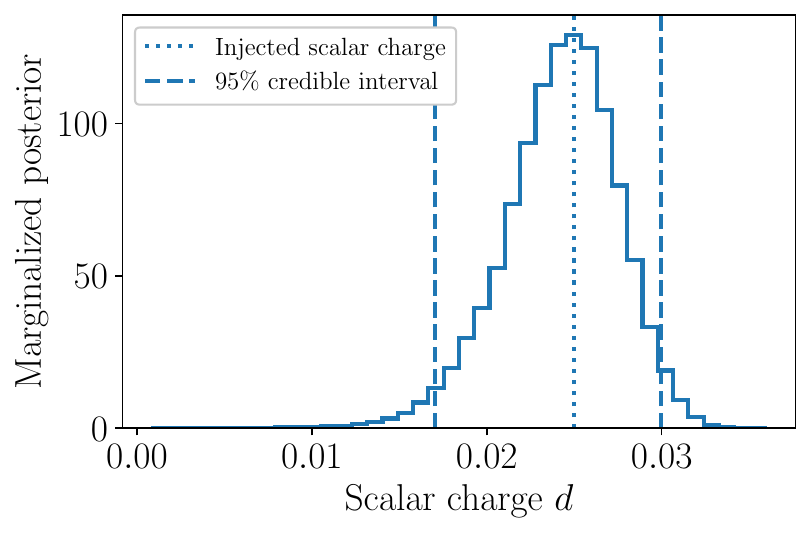}
    \caption{Marginalized posterior distribution of an EMRI system with scalar charge $d=0.025$ and source parameters $M=10^5\,M_\odot$, $\mu=5\,M_\odot$, $a=0.95$, $e_0=0.4$, $T=2$ yrs and SNR=50.
    The estimated median and 95\% credible interval are $0.0244^{+0.006} _{-0.007}$.
    }
\label{fig:charge_histogram}
\end{figure}

For the same system, we also explore the impact of 
ignoring the scalar charge and fitting the data with 
a GR template (see Appendix~\ref{app:bias} for further details). 
We find that the GR waveform recovers the injected signal 
with $2-3\sigma $ biases in the source intrinsic parameters, 
i.e., its masses and spins. While such systematic errors are 
large compared to the size of the posterior, they can be 
considered small for astrophysically motivated 
studies. {For instance the central black hole {dimensionless} spin $a$ estimated with a GR template is shifted by $\approx 4\times 10^{-5}$ from the injected value $a=0.95$.}


To illustrate how a bound in $d$ can be converted to a constraint on a specific theory, we now consider GB gravity defined by the action in Eq.~\eqref{eq:actionGB}. The forecasted constraints on the coupling constant are shown in Fig.~\ref{fig:alpha_bound}. An interesting feature is that the strongest bound for $\sqrt{\alpha}$ comes from system 2, while the strongest bound for $d$ came from system 1. This is because the relation between $d$ and $\alpha$ involves the mass of the secondary. 

Selecting a specific theory also allows for a comparison between bounds from EMRIs and bounds from other systems. 
The analysis of the event GW230529 \cite{GW230529,Gao:2024rel,TGR_GW230529} yielded
$\sqrt{\alpha}_{95\%}=1.4 \, \rm{km}$, which is a few percent larger than the EMRI constraint obtained with systems 1, 2, and 3. Interestingly, the forecasted best constraint on $\sqrt{\alpha}$ for LVK Voyager is larger (see extremal bound from Figure~21 of \cite{Perkins:2020tra}).
The initial observational frequency of GW230529 is {$f=$}20 Hz{$\approx 10^{-4}M_\odot$}, and the total system mass is {$M_{\rm tot}=$}$5.1 \, M_\odot$ and its SNR=11.1~\cite{GW230529}. The dimensionless velocity of GW230529 is $v=(\pi M_{\rm tot} f)^{1/3} \approx 0.117$, whereas the same {initial} velocity for the lowest total mass EMRI system, $10^5\,M_\odot + 5\,M_\odot$, is $v=(\Omega_\phi)^{1/3} \approx 0.23$ {(see Appendix~\ref{app:pn_bounds} for details on the velocity definitions)}. This means that GW230529 is in a weaker field compared to the EMRI systems we considered. 
{In fact, since scalar emission becomes more significant at lower frequencies due to its 
leading dipole character, we expect that considering non-plunging EMRIs with an initial 
dimensionless velocity of $v \approx 0.1$ and $p_0 \approx 60 $ would further strengthen 
EMRI constraints. For the same reason, intermediate-mass ratio inspirals with sufficiently 
small secondaries could provide even tighter constraints on fundamental fields. This is because 
their observed inspiral can begin in much weaker field regions, \textit{i.e.}, at larger orbital 
separations.}

\begin{figure}
\centering
\includegraphics[width=\columnwidth]{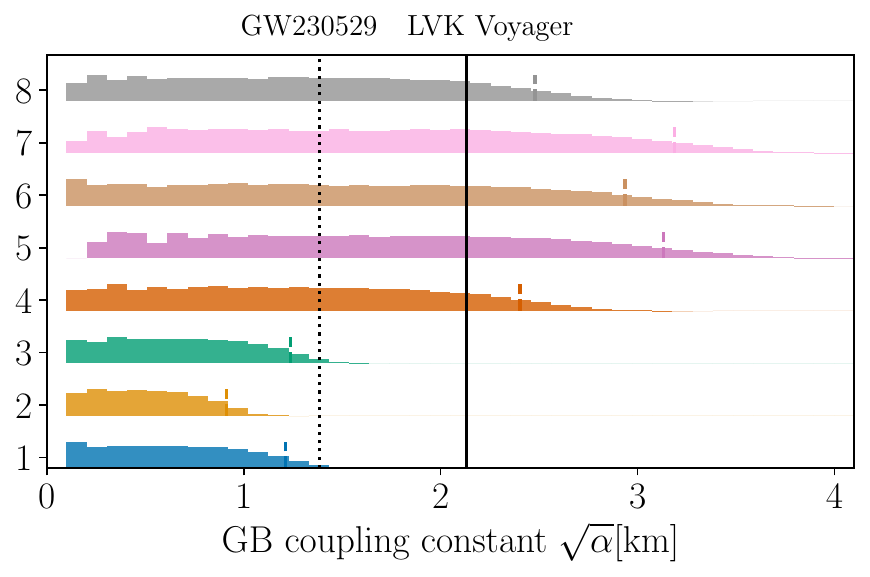}
        \caption{Posterior distribution of the 
        Gauss-Bonnet coupling mapped from the scalar charge constraints (see Fig.~\ref{fig:charge_bound}).
         The black dotted line in the right panel shows the bound \cite{Gao:2024rel} from the observation of the gravitational wave event GW230529 \cite{GW230529}, while the black solid line shows the best forecasted bound for LVK Voyager configuration obtained in \cite{Perkins:2020tra}. 
         We use a different normalization with respect to \cite{Gao:2024rel,Perkins:2020tra} (see Appendix~\ref{app:normalization_action}).}\label{fig:alpha_bound}
    \end{figure}
\noindent{{\bf{Discussion.}} 
In this work, we produced the first ready-to-use 
fully relativistic EMRI waveforms in theories 
of gravity that include a massless scalar 
field and perform the first Bayesian analysis 
of EMRI signals in this context. The phase of our waveforms is modeled at 
the adiabatic order {within the flux accuracy} and includes a single extra 
parameter, the scalar charge of the EMRI 
secondary. This is the only parameter needed to capture the effect of the scalar field at this order \cite{Maselli:2020zgv}.

Our study provides Bayesian methods and gravitational wave models to investigate the fundamental physics potential of {single EMRI systems in future} LISA observations. These are two key objectives outlined in the LISA definition study report \cite{Colpi:2024xhw} and the fundamental physics white paper \cite{LISA:2022kgy}. We also provide an approximate comparison of EMRI constraints on agnostic PN deviations in the phase in Appendix~\ref{app:pn_bounds}, where we find that EMRIs provide three orders of magnitude improvement compared to current detectors.

{We forecast the expected bound on the scalar charge of the secondary, $d$, in a theory-agnostic manner.}
Bounds on $d$ can be translated to bounds on the coupling constant(s) or a particular theory. We have demonstrated this for a specific example, that of linear-Gauss-Bonnet gravity. Focusing on a specific theory allowed us to compare our forecast with existing bounds and forecasts for the next generation of GW detectors for this particular theory. 
One of the caveats of this comparison is that our exploration of the parameter space is not exhaustive. As such, it is not clear if the EMRI parameters we have considered are the ones that would yield the most stringent bounds for either $d$ or for the coupling constant of some particular theory. We plan to address this question in future work. 

{There are several ways in which this work can be extended.}
Our waveform model can be extended in several 
directions. Realistic EMRIs are expected to follow 
generic, inclined orbits \cite{DellaRocca:2024pnm}. 
Studies of EMRIs with new fundamental fields 
evolving on eccentric \textit{and} inclined 
orbits are underway \cite{papergeneric}. Inclusion of post-adiabatic corrections, along the lines of Ref.~\cite{Spiers:2023cva}, is also essential to include known GR effects that enter at post-adiabatic order, such as the secondary spin \cite{Burke:2023lno}, and to assess how new fundamental physics can affect the waveform at this order. 
Our approach can also be generalised to capture massive scalars \cite{Barsanti:2022vvl} and more general vector/tensor fields \cite{Fell:2023mtf}. 
Finally, the presence of environmental effects in realistic EMRI signals might affect the detectability of the scalar charge through parameter correlations \cite{Kejriwal:2023djc}, so further joint studies are required. {Future studies will need to incorporate generic waveforms and more comprehensive modeling of environmental and beyond-GR effects.}

{Our Bayesian analysis could be extended to account for the presence of multiple sources, as expected in the LISA data stream. Future gravitational wave inference pipelines will need to explore how tests of fundamental physics can be performed within a global fit framework \cite{Strub:2024kbe,Littenberg:2023xpl,Deng:2025wgk,Katz:2024oqg}. In particular, it will be important to assess how the presence of additional sources affects the constraints obtained in single-source scenarios such as the one presented here. More importantly, these pipelines must evaluate how the detectability of deviations from General Relativity may be influenced by the global fit procedure.}

\noindent{{\bf{Acknowledgments.}}}
We thank Scott Hughes for providing flux values for comparison with our implementation and  Gabriel Andres Piovano for implementing the homogeneous solutions' functions for the gravitational fluxes computation \cite{Piovano:2021iwv}. We thank L. Pompili, E. Sänger, and A. Buonanno for their comments on the manuscript. This work makes use of the Black Hole Perturbation Toolkit \cite{BHPToolkit}. In particular, the \texttt{KerrGeodesics} \cite{KerrGeodesics} and \texttt{SpinWeightedSpheroidalHarmonics} \cite{SpinWeightedSpheroidalHarmonics} were employed. Some numerical computations have been performed at the Vera cluster supported by MUR and by Sapienza University of Rome.
A.~Maselli acknowledges financial support from MUR PRIN Grant 
No.~2022-Z9X4XS, funded by the European Union - Next Generation EU.
T.P.S. acknowledges partial support from
the STFC Consolidated Grants no.~ST/X000672/1 and
no.~ST/V005596/1. 
N.~Warburton acknowledges support from a Royal
Society - Science Foundation Ireland University Research Fellowship. This publication has emanated from research conducted with the financial support of Science Foundation Ireland under grant number 22/RS-URF-R/3825.
M.~van de Meent acknowledges financial support by the VILLUM Foundation (grant no. VIL37766) and the DNRF Chair program (grant no. DNRF162) by the Danish National Research Foundation. 
O.~Burke acknowledges support from the French space agency CNES in the framework of LISA.

\noindent{{\bf{Author Contributions.}}}
The Author Contributions statement describes the contributions of individual authors referred to by their initials, and, in doing so, all authors agree to be accountable for the content of the work (see \cite{credit_contribution} for more information and \href{https://www.cell.com/pb/assets/raw/shared/guidelines/CRediT-taxonomy.pdf}{here} for the taxonomy of contributor roles).

LS: Conceptualization, Data Curation, Formal Analysis, Investigation, Methodology, Project Administration, Resources, Software, Supervision, Validation, Visualization, Writing – Original Draft.

SB: Data Curation, Investigation, Resources,  Software,  Writing - Review \& Editing

AM: Conceptualization, Methodology, Writing - Original Draft, Writing - Review \& Editing, Supervision, Project Administration. 

TPS: Conceptualization, Methodology, Writing - Original Draft, Writing - Review \& Editing, Supervision, Project Administration. 

NW: Data Curation, Software, Writing - Review \& Editing

MvdM: Methodology, Software, Writing - Review \& Editing.

AJKC: Conceptualization, Methodology, Writing - Review \& Editing.

OB: Investigation, cross-validation (MCMC \& FM studies), Data curation, Writing - Review \& Editing 

JG: Methodology, Writing - Review \& Editing, Supervision.

\appendix

\section{Normalization of the action}\label{app:normalization_action}
We provide here details on the mapping between 
constraints on the Gauss-Bonnet coupling parameter inferred in this paper and from current and future ground-based detectors.

To compare bounds on the Gauss-Bonnet 
coupling $\alpha$ we need to take into account the 
different normalizations considered in the literature for 
the actions \eqref{eq:action} and \eqref{eq:actionGB}. 
In particular, constraints derived from current GW 
observations \cite{Lyu:2022gdr} and by the network of future detectors \cite{Perkins:2020tra} are obtained by 
assuming the following action for GB gravity:
\begin{equation}
S = \int d^4 x\sqrt{-g}\left[\frac{R}{\kappa}-\frac{1}{2} (\nabla \bar{\phi})^2  + \bar{\alpha}_{\rm GB} \bar{\phi} \mathcal{G}
     \right] \ , \label{eq:actionthere}
\end{equation}
where $\kappa=16\pi${, and the scalar field $\bar{\phi}$ and constant $\bar{\alpha}_{\rm GB}$ are related to the action considered in this work}
\begin{equation}
S = \int d^4 x\frac{\sqrt{-g}}{\kappa}\left[R-\frac{1}{2} (\nabla \phi)^2  + \frac{\alpha}{4} \phi \mathcal{G}
     \right] \ ,\label{eq:actionhere}
\end{equation}
{via the relations $\bar{\phi}= \phi \kappa^{-1/2}$ and $\alpha= 4\kappa^{1/2} \bar{\alpha}_{\rm GB}$. Using this last equation we provide the conversion of the bounds presented in the literature to our conventions.} The constraint available for GB gravity of \cite{Lyu:2022gdr}, $\sqrt{\bar{\alpha}_{\rm GB}}\simeq 1.18$ km, translates to $\sqrt{\alpha}=4\pi^{1/4}\sqrt{\bar{\alpha}_{\rm GB}}\simeq 6.3$ km and the constraints from GW230529 $\sqrt{\bar{\alpha}_\tn{GB}} \simeq 0.260 \tn{km} \rightarrow \sqrt{\alpha_\tn{GB}} \simeq 1.4 \tn{km}$ \cite{Gao:2024rel}. 
Similarly, for the projected bound by LIGO O8, $\sqrt{\bar{\alpha}_{\rm GB}}\simeq 0.4$km (see Fig.~21 of \cite{Perkins:2020tra}), 
we obtain$\sqrt{\alpha}\simeq 2.1$km.

The map between the scalar charge and the GB 
coupling has been obtained in \cite{Julie:2019sab}, 
using the action 
\begin{equation}
    S_{J} = \int \frac{\sqrt{-g}}{\kappa}d^4x \left[R-2(\nabla \tilde{\phi})^2  + 
    \tilde{\alpha}\tilde{\phi}\mathcal{G}\right] \ .
\end{equation}
In this setup, at the leading order in 
$\tilde{\alpha}$, the scalar charge is given 
by $\tilde{d}=\tilde{\alpha}/(2\mu^2)$. 
Passing to the {normalization} we use in Eq.~\eqref{eq:actionhere}, 
$\tilde{d}=d/2$ and $\tilde{\alpha}=\alpha/2$, 
which leaves the relation $d=d(\alpha)$ 
unchanged.

\section{Flux interpolation}\label{app:interpolation}

Energy and angular momentum fluxes are interpolated over a three-dimensional grid constructed using 13 Chebyshev-Gauss-Lobatto (CGL) nodes for the eccentricity $e\in[0.0, 0.5]$ and the primary {dimensionless} spin $a\in[-0.99,0.99]$. Rather than using the semi-latus rectum, we find it more convenient to introduce the variable
\begin{equation}
    u = (1+e)\qty[\frac{\Omega_\phi (a, p, e)}{\Omega_\phi \left(a, \frac{p_{\rm sep}(a,e)}{1+e}, e\right)}]^{2/3}\ ,
\end{equation}
where $p_{\rm sep}(a,e)$ is the separatrix for Kerr spacetime, and negative spins correspond to retrograde orbits. We compute $u$ on 17 CGL nodes within $u \in [0.08,0.97]$. Therefore, we calculate scalar and gravitational fluxes on a total of $13\times 13\times 17 = 2873$ grid points. 

Before interpolation, we normalize the fluxes by their leading order contribution in a post-Newtonian expansion. We construct 4 Chebyshev interpolants for the gravitational and scalar energy and angular momentum fluxes. 
Assuming the fluxes are smooth functions of the variables on our interpolation domain, the accuracy of the Chebyshev interpolation should converge exponentially with the number of grid points. Consequently, we can estimate the interpolation ``aliasing'' error from the magnitude of the coefficients of the highest order Chebyshev polynomials in each direction.
The interpolated quantities {are the dimensionless scaled fluxes $\dot{\tilde{E}} = \dot{E}/q^2, \dot{L}_r = \dot{\tilde{L}}/(q^2 M)$} and their Chebyshev errors $\sigma $ are shown in Table~\ref{tab:cheby_int}.

If we write the total energy flux as:
\begin{equation}
    \dot{\tilde{E}} = \frac{32}{5 p^5}\qty(\frac{5p^{5}}{32}\dot{\tilde{E}}_{\rm grav} + \frac{5p^{5}}{32} {d^2}\dot{\tilde{E}}_{\rm scal})\ ,
\end{equation}
we can estimate the size of the scalar charge that is comparable to the error in the gravitational fluxes
\begin{equation}
     \frac{5p^5}{32} \dot{\tilde{E}}_{\rm scal} d^2> \sigma_{\rm grav}\rightarrow
     0.12 \, d^2>\sigma_{\rm grav} \, ,
\end{equation}
where we inserted $p^4 \, \dot{\tilde{E}}_{\rm scal} =0.3$ which is a typical value accross the grid for $p=10$. This gives a relation between the scalar charge and the Chebyshev interpolation. For values of $d=0.01$, we obtain a constraint on the error of the order $10^{-5}$, which is one order of magnitude smaller than what we obtained with our interpolation scheme.  This does not invalidate our work, as we treat our interpolated fluxes as the true fluxes and use them consistently for injection and recovery. However, this does highlight the need for denser flux grids and accurate interpolation schemes. The production of dense self-force grids with accurate interpolation methods is a current major challenge for extending the FastEMRIWaveforms package to fully generic orbits in Kerr spacetimes. 
  
Using the \href{https://github.com/BlackHolePerturbationToolkit/GremlinEq?tab=readme-ov-file}{Gremlin code} for flux calculations evaluated at $a=0.95$, $ p=10.1930405906075$, $e=0.4081632653061225$ we obtain a flux value for $\frac{5}{32} p^5\dot{\tilde{E}}_{\text{grav}}$ which differs from our interpolation by absolute and relative errors of $3.8\times 10^{-4},3.6\times 10^{-4}$, respectively. This is compatible within our estimated interpolation error. 
{Given the recent work \cite{Chapman-Bird:2025xtd}, we compared the fluxes at $p=10.0,e=0.4,a=0.5$ and found a relative difference of $7\times 10^{-5}$ which can be compared to Figure~5 of \cite{Chapman-Bird:2025xtd}.
}
\begin{table}[htbp!]
\centering
\caption{
Chebyshev interpolation of the energy and 
angular momentum fluxes and the estimates of their absolute interpolation errors.
}
\vspace{0.2cm}
\begin{tabular}{c|c}
\hline
\hline
{Interpolated expression} & Abs. error \\ 
\hline
$\frac{5}{32} p^5\dot{\tilde{E}}_{\text{grav}}$ & $7.5\times 10^{-4} $ \\ 
$\frac{5}{32} p^{7/2}\dot{\tilde{L}}_{\text{grav}}$ & $7.1\times 10^{-4} $ \\ 
$p^4 \dot{\tilde{E}}_{\text{scal}}$ & $2.3\times 10^{-4} $ \\ 
$p^{5/2} \dot{\tilde{L}}_{\text{scal}}$ & $2.2\times 10^{-4} $ \\
\hline
\hline
\end{tabular}
\label{tab:cheby_int}
\end{table}

Having the fluxes in hand, we can obtain the right hand side of the Eqns.~\eqref{eq:EMRIODE} and then use the FEW package to obtain the EMRI trajectory. 
As an example, we show in Fig.~\ref{fig:trajectory} the trajectories in the $p-e$ plane for binaries with $a=0.95$ in GR, i.e., setting the scalar charge to zero, and different EMRI masses. For reference we also show the Chebyshev grid points at the $a=0.9562665680261776$.
\begin{figure}[hbpt!]
    \includegraphics[width=\columnwidth]{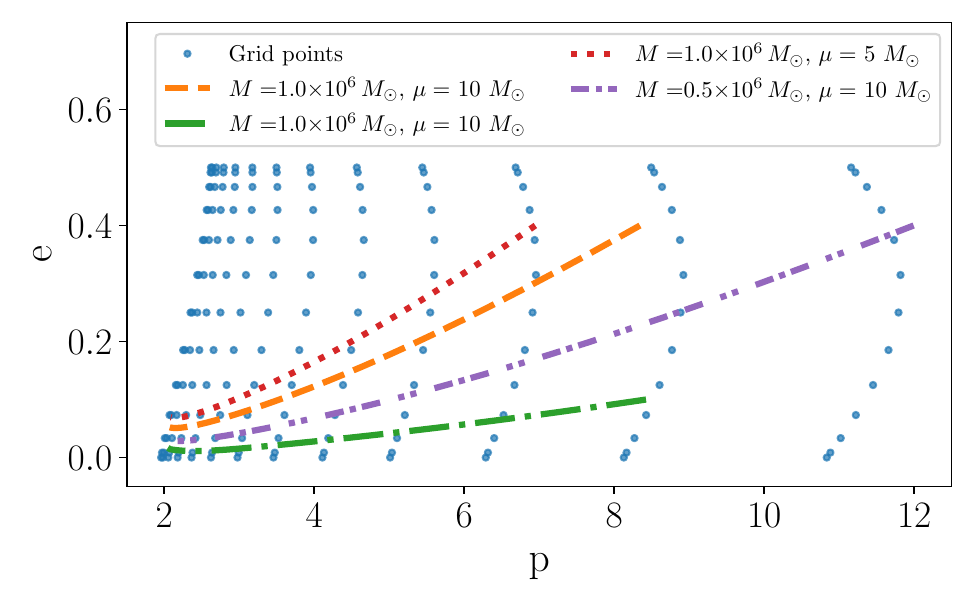}
    \centering
    \caption{ 
    Trajectory evolution of semi-latus rectum and eccentricity for four EMRIs with different component masses and $a=0.95$. For reference, we show the location of the grid points used for interpolation in the $(p,e)$ plane for a constant {dimensionless} spin slice at $a=0.9562665680261776$, where the largest value of $p$ reached for this {dimensionless} spin is approximately $p\approx 30$.}
    \label{fig:trajectory}
\end{figure}

\section{Data analysis setup}\label{app:datanalysis}
The posterior distributions presented in this work are obtained using MCMC sampling with the package \texttt{Eryn} \cite{Karnesis:2023ras, emcee}. To run the MCMC, we need to specify the priors $p(\Theta)$ and the likelihood function
\begin{equation}
    \ln p(s|\Theta) =-\frac{1}{2}\braket{s-h(\Theta)}{s-h(\Theta)} \, ,
\end{equation}
where we defined the inner product between two GW templates $h_1(t)$ and $h_2(t)$ \cite{Speri:2022kaq}:
\begin{equation}
\braket{h_1(t)}{h_2(t)} = 4 \Re \int_0 ^\infty \frac{\tilde h_1^*(f) \tilde h_2(f)}{S_n(f)} \, \dd f\ .
\end{equation}
The noise spectral density $S_n(f)$ of LISA is taken from \cite{Babak:2021mhe} and assumed to be known. The tilde denotes the Fourier transform of the waveform and the symbol $^*$ denotes complex conjugation. The Fourier transform and likelihood evaluation are performed on GPUs using \texttt{cupy}~\cite{cupy}.
Before passing to the frequency space, we taper $h(t)$ with a Tukey window. 
The parameter that controls the magnitude of the sinusoidal lobes of the window has been fixed to \texttt{alpha}$=0.005$ \cite{scipy}. The sampling interval was adjusted for different black hole masses $M$ to avoid aliasing.

The full parameter space of the AAK model is given in Table~\ref{tab:priors}. The response of the detector to the  
signal is described by the two data channels $h_I(t;\Theta)$ and $h_{II} (t;\Theta)$. The full log-likelihood is given by the sum of the log-likelihood for each channel.
For reference, we show in Figure \ref{fig:spectrogram} the spectrogram of $h_I$ for a system with parameters $M=10^6 \, M_\odot, \mu=10\,M_\odot,e_0=0.4,d=0.0025,p_0=8.3 ,T=2\,{\rm yrs}$. 
\begin{figure}[hbpt!]
    \centering
\includegraphics[width=\columnwidth]{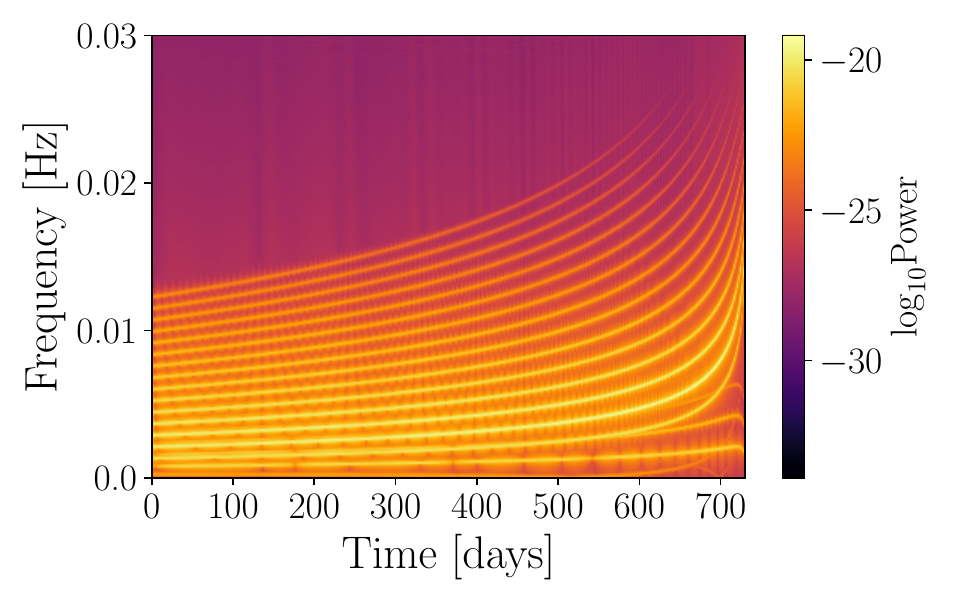}
    \caption{Spectrogram of the gravitational wave signal output of the AAK waveform obtained from an EMRI system with parameters $M=10^6 \, M_\odot, \mu=10\,M_\odot,e_0=0.4,d=0.0025,p_0=8.3 ,T=2\,{\rm yrs}$. 
    The different color bands represent the different harmonics, and their color intensity represents their power.}
    \label{fig:spectrogram}
\end{figure}

We assume flat priors for all parameters apart from the luminosity distance, which is assumed to follow a power-law distribution with slope $-2$ in the range $[0.01,10.0]$~Gpc.
A summary of the priors we consider is given in Table~\ref{tab:priors}. 
We restrict the prior in $\Phi_{\phi_0}$ to $\pi$ and not to $2 \pi$ because there is an exact degeneracy every $\pi$. 
The prior choice of power-law with slope $-2$ for the luminosity distance is motivated by the fact that we found some chains getting stuck at large unphysical values of $d_L$. This was caused by chains getting stuck on secondary modes of the likelihood. 
Our choices allow for better sampling efficiency without affecting the results.
\begin{table}[htbp!]
\caption{Prior distributions on the waveform parameters 
used for MCMC posterior sampling.}
\centering
\begin{tabular}{c|c}
\hline
\hline
{Parameter} & {Priors ($\delta=0.01$)} \\ 
\hline
\hline
$\ln M/M_\odot$ & Uniform $[\ln M^* (1- \delta), \ln M^* (1+ \delta)]$ \\ \hline
$\ln \mu /M_\odot$ & Uniform $[\ln \mu^* (1- \delta), \ln \mu^* (1+ \delta)]$ \\ \hline
$a$ & Uniform $[a^* (1- \delta), 0.98]$ \\ \hline
$p_0 $ & Uniform $[p_0 ^* (1- \delta), p_0 ^* (1+ \delta)]$ \\ \hline
$e_0$ & Uniform $[e_0 ^* (1- \delta), e_0 ^* (1+ \delta)]$ \\ \hline
$d_L$ [Gpc] & Power Law $[0.01, 10.0]$ \\ \hline
$\cos \theta_S$ & Uniform $[-0.99999, 0.99999]$ \\ \hline
$\phi_S$ & Uniform $[0.0, 2 \pi]$ \\ \hline
$\cos \theta_K$ & Uniform $[-0.99999, 0.99999]$ \\ \hline
$\phi_K$ & Uniform $[0.0, 2 \pi]$ \\ \hline
$\Phi_{\phi_0}$ & Uniform $[0.0, \pi]$ \\ \hline
$\Phi_{r_0}$ & Uniform $[0.0, 2 \pi]$ \\ \hline
$\Lambda$ & Uniform $[-0.6, 0.6]$ \\ 
\hline
\hline
\end{tabular}
\label{tab:priors}
\end{table}
For parameters that are typically constrained with high precision, i.e., $(M,\mu,a,p_0,e_0)$, we center the priors around the true injected values $\Theta_{\rm true}$. In all runs, we find that the posterior support of all parameters is much tighter than the assumed prior.

{Here, we provide a motivation for sampling in $\Lambda = d^2$.
The deviation due to the scalar emission enters the fluxes and the dephasing from vacuum as $d^2$ (see Figure~\ref{fig:phase_difference} and discussion in Appendix~\ref{app:ode_error}). This means that the waveform difference between a scalar charge template and a vacuum one is approximately $h(d\neq 0) -h(d=0) \approx h(d=0)e^{i \Delta \Phi}  $ where the dephasing $\Delta \Phi$ is a small number proportional $d^2$. Therefore, the log-likelihood becomes approximately Gaussian in $d^2$.}
For this reason,
we sample in the parameter $\Lambda = d^2$ with a uniform prior ${\cal U}_{[-0.6,0.6]}$, and then select the samples with $\Lambda>0$. Sampling in $\Lambda$, both positive and negative, allows us to obtain {near-}Gaussian posteriors, which can be sampled more efficiently with MCMC methods{ and the chosen proposals. We also verified that sampling $\Lambda$ with a uniform positive prior would not change the posterior. However, we found that this would require running the MCMC for longer to accumulate a larger number of effective samples}
We use two proposals to efficiently sample: the stretch move \cite{emcee} and an adaptive metropolis move that jumps along the eigen-directions of the covariance matrix. We sample the posteriors using 26 walkers, monitoring the integrated autocorrelation time $\tau$ as a function of the iteration. We assume that the estimator for $\tau$ is reliable when it plateaus below the line given by $N_{\text{it}}/50$ for $N_{\text{it}}$ the number of iterations (see \href{https://emcee.readthedocs.io/en/stable/tutorials/autocorr/}{this link} for further details).
We show in the upper corner plot of Figure~\ref{fig:full_posterior} the full posterior distribution used for the constraints obtained in Figures~\ref{fig:charge_bound} and \ref{fig:alpha_bound}.

\begin{figure*}
    \centering
    \includegraphics[width=\textwidth]{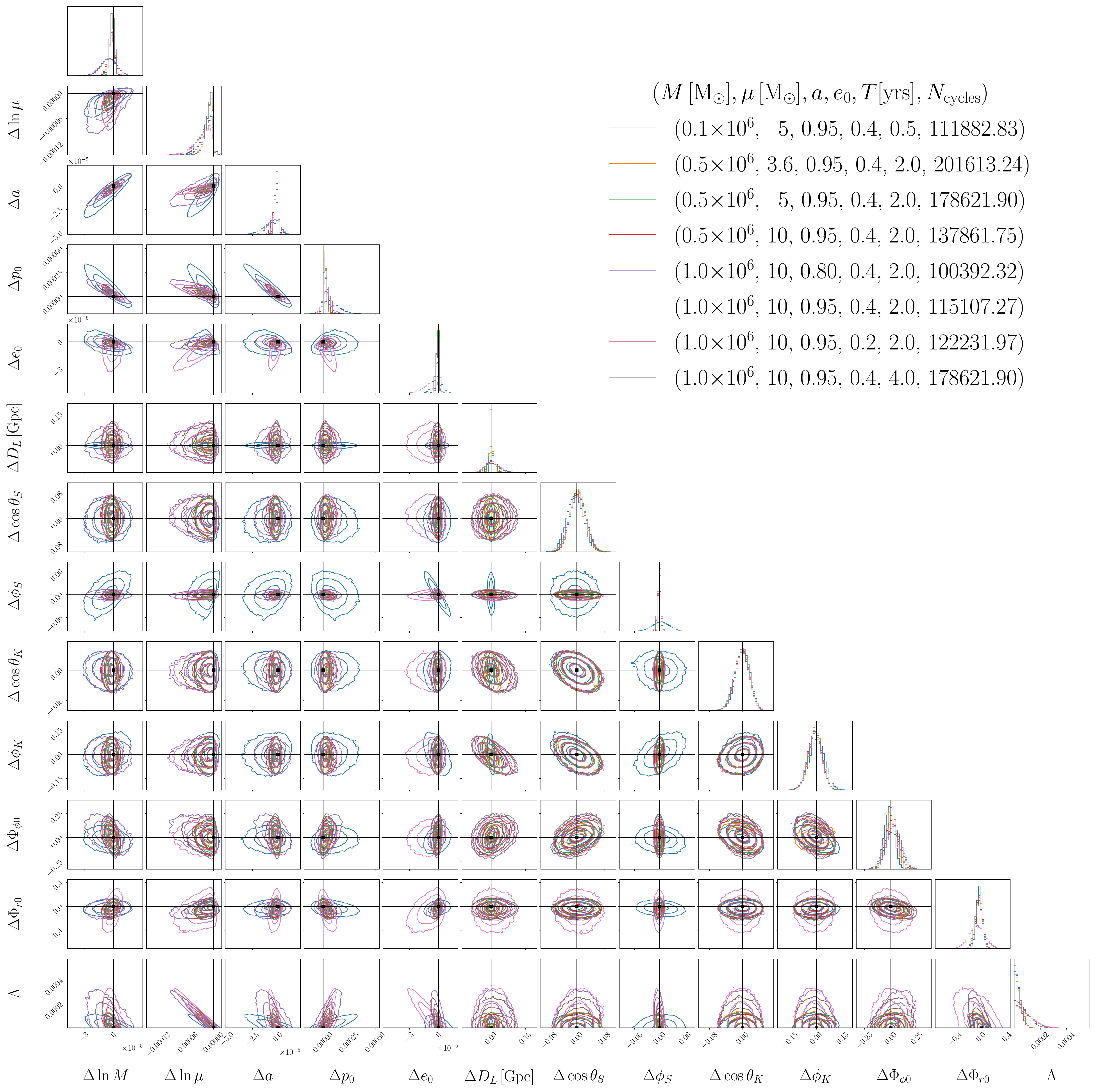}
    \includegraphics[width=\textwidth]{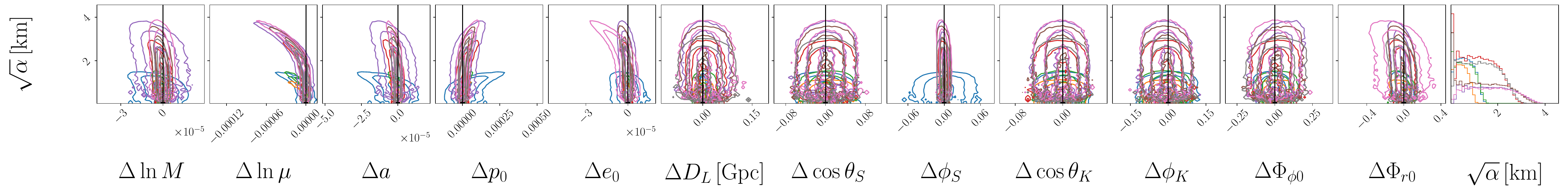}
    \caption{
    Posterior distribution of EMRI injections with different orbital configurations. The posteriors are centered around the injected parameters.
    Diagonal and off-diagonal plots provide marginalised and 2D-joint posteriors, respectively. Contour lines in off-diagonal panels identify the
    1,2,3$-\sigma$ Gaussian credible contours of each distribution.
    The upper corner plot shows the posterior distribution output of the MCMC analysis in the parameters of Table~\ref{tab:priors}. The lower corner shows how the list row of the upper corner plot transforms when mapping to the coupling $\sqrt{\alpha}$.
    In the legend, we provide the system parameters and the number of orbital cycles performed $N_{\rm cycles}$.
    {A downloadable version of these plots can be found \href{https://github.com/lorenzsp/testGRwEMRIs/tree/main/plot_paper/figures}{here}.}
    }
    \label{fig:full_posterior}
\end{figure*}
To obtain a bound on this specific theory from the posteriors on $\Lambda = d^2 $ and $\ln \mu$, we use
\begin{align}
\sqrt{\alpha} = \sqrt{2} \mu \sqrt{d} = \sqrt{2} \, \mu  \, \Lambda^{1/4} \, ,
\end{align}
and the determinant of the Jacobian of such transformation 
\begin{equation}
\dd \sqrt{\alpha} = \qty|\pdv{\sqrt{\alpha}}{\Lambda}| \dd \Lambda \propto  \mu  \, \Lambda^{-3/4} \dd \Lambda\ .
\end{equation}
Similarly, one can obtain the bound on the scalar charge $d$.
In the lower corner plot of Figure~\ref{fig:full_posterior}, we can see how the posterior samples are mapped to the coupling $\sqrt{\alpha}$. The posteriors are non-Gaussian and have long tails. {This is not only due to the quartic root relation between $\sqrt{\alpha}$ and $\Lambda$, but also due to the strong correlations of $\Lambda$ with the secondary mass $\mu$. In fact, the deviations from vacuum in the fluxes and phases are both proportional to $\mu$ and $\Lambda$.} This demonstrates the importance of sampling in $\Lambda$ instead of $\sqrt{\alpha}$. Sampling in the latter requires longer iterations to reach convergence and to resolve the tails of the ``banana-shaped'' distributions. 

\section{Mapping to agnostic bounds}\label{app:pn_bounds}
To compare the potential of EMRIs to constrain deviations from General Relativity for agnostic parametrization, we provide a comparison in terms of the parametrized post-Einsteinian (ppE) formalism or Flexible Theory Independent formalism \cite{Ajith:2024mie,Mishra:2010tp,Yunes:2009ke,Gupta:2024gun,Li:2011cg,Agathos:2013upa,Yunes:2016jcc,Mehta:2022pcn,Yunes:2010qb}.
We consider the system with $10^6\,M_\odot+10\,M_\odot$ solar masses, initial eccentricity $e_0=0.4$, and time to plunge $T=2$ years and run a GR inspiral. We obtain the gravitational energy fluxes $\dot{E}_{\rm grav}$ from the samples obtained, evaluated at the start of the inspiral. If we subtract the median value from this set of samples, and divide by the median, we obtain a set of samples representing the fractional deviation from the expected energy dissipation in GR, that are consistent with the posterior. These can be related to ppE deviations by writing
\begin{equation}
    \Delta \dot{E}/\dot{E}_{\rm grav} = B v^{2n}
\end{equation}
where $v= \qty(\pi \qty(M + \mu) f)^{1/3}$ with $f$ frequency of the $(l,n_\phi,n_r)=(2,2,0)$ and $n$ the 
post-Newtonian (PN) order. For EMRIs we can approximate 
$v = \Omega_\phi ^{1/3}$. 
The quantity $B$ can be mapped to an ``agnostic'' deviation in the waveform phase $\delta \varphi$ at different PN orders using 
the formalism described in \cite{Yunes:2016jcc} 
(see Eqns.~(9)-(11) and (19)-(28) \cite{Yunes:2016jcc} 
for the mapping $B\rightarrow\delta\varphi$).
We {provide} constraints on $\delta\varphi$ obtained 
with this procedure as ``GR mapping'' 
in Fig.~\ref{fig:bound_delta_phi}. Constraints are $95\%$ upper limits obtained from the posterior on $\delta \varphi$.  

\begin{figure}[hbpt!]
    \centering
    \includegraphics[width=\columnwidth]{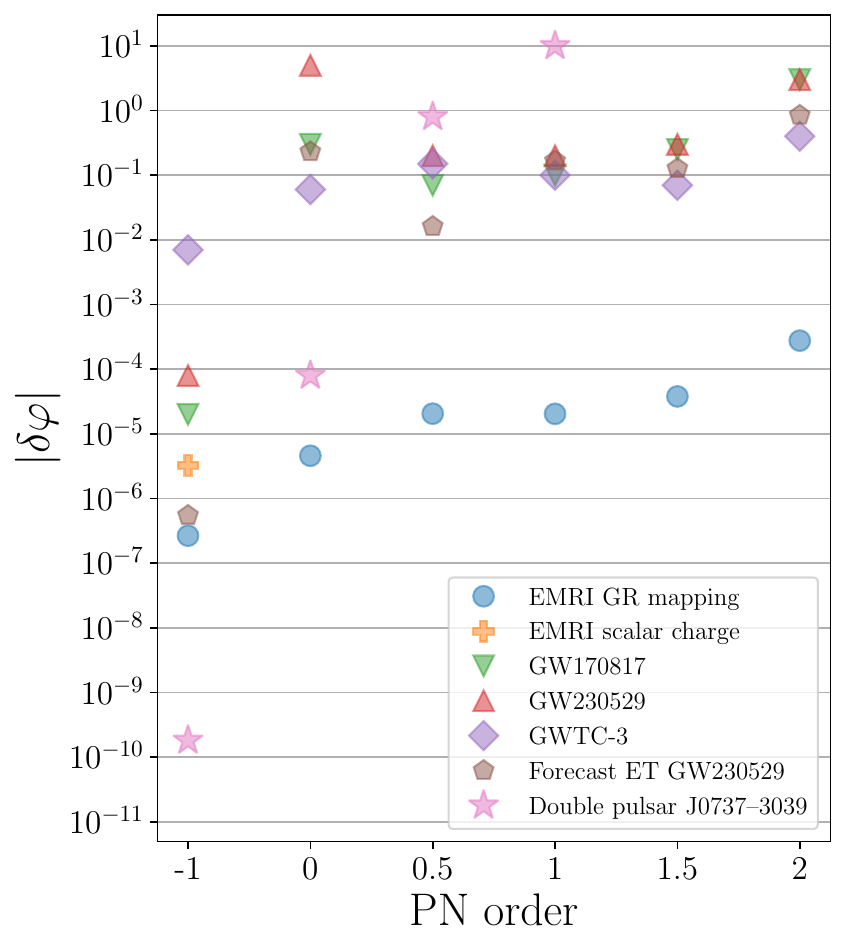}
    \caption{Comparison of the constraints on the phase deviation at different PN orders. The EMRI constraints are obtained from mapping the posterior distribution of a system with parameters $M=10^6 \, M_\odot, \mu=10 \, M_\odot, a=0.95, e_0=0.4, T=2 {\rm yrs}$ and SNR=50 into the phase deviation at different PN orders (blue dots, EMRI GR mapping). For the case of $-1$PN order we use the posterior distribution obtained from an EMRI embedded in a scalar field (orange cross) as done in Figure~\ref{fig:charge_bound}. We show the current constraints obtained from the gravitational wave events GW170817 (down green triangle) \cite{TGR_GW170817}, GW230529 (up red triangle) \cite{TGR_GW230529}, and using the GW transient catalogs of the third observing run (GWTC-3) 
    (violet rombo) \cite{LIGOScientific:2021sio}. 
    Brown pentagon markers correspond to forecasts 
    on $\delta\varphi$ obtained for a 
    GW230529-like binary observed by ET (see 
    main text). The double pulsar constraints obtained from PSR J0737–3039 \cite{Kramer:2021jcw} are shown with 
    pink star markers.}
    \label{fig:bound_delta_phi}
\end{figure}

We remark that this analysis is approximate for three reasons.
Firstly, post-Newtonian expansions do not provide 
a good description of the evolution of EMRIs. Secondly, this 
mapping only considers deviations in the waveforms that can be described by changes in the GR parameters. While any GR deviation causing such a change would definitely not be detectable, larger deviations could also be undetectable, since they are clearly strongly correlated with changes in the EMRI parameters. In that sense, this should be considered an optimistic bound. We note that this method cannot assess the detectability of components of the deviation that change the waveforms in ways that are orthogonal to the GR waveform space, but we expect these to be sub-dominant.
{Thirdly, we are mapping an eccentric inspiral to a quasi-circular ppE parametrization.
}

For comparison, we also provide the mapping between 
the agnostic approach and the scalar 
charge at the -1PN order, which would correspond 
to the leading contribution of our 
full adiabatic scalar emission. In this case, 
the parameter $B$ is given by \cite{Barausse:2016eii}:
\begin{equation}
    B= d^2  \, \Omega_\phi ^{2/3} \dot{E}_{\rm scal}/\dot{E}_{\rm grav} \, .
\end{equation}

In Fig.~\ref{fig:bound_delta_phi}, we show the constraints obtained using the GR mapping at different PN orders for the EMRI configuration we considered (blue dots). The scalar charge mapping shows degradation of one order of magnitude (orange cross) at the -1PN order. This demonstrates the importance of correlations that are not taken into account in the GR mapping. 

The constraints obtained from the gravitational wave events GW170817 \cite{TGR_GW170817}, GW230529 \cite{TGR_GW230529}, and using the GW transient catalogs of the third observing run (GWTC-3) \cite{LIGOScientific:2021sio} are also shown, and these are a few orders of magnitude larger than the EMRI constraints. This is expected since EMRIs complete $10^4-10^5$ cycles during the observation, which is two to three orders of magnitude than the number of cycles typically observed in a merger seen by ground-based detectors, leading to correspondingly higher measurement precision. However, better phase constraints do not necessarily imply tighter bounds on the coupling, as this map depends on the specific theory considered. 
This is the case for GB gravity for which constraints improve for lighter objects, as in the case of GW230529 (see vertical line Fig.~\ref{fig:alpha_bound}).

For comparison, we compute bounds 
on $\delta\varphi$ forecasted for a third-generation ground-based detector like the Einstein 
Telescope (ET). Constraints are derived 
through a Fisher matrix approach, 
analysing the inspiral phase of a binary 
BH system with the same properties as GW230529 
(the source parameters are fixed to the 
median values reported in \cite{LIGOScientific:2024elc}). 
For the analysis, we consider a TaylorF2 
waveform model, 
integrated between 3Hz 
and the Schwarzschild ISCO\footnote{Constraints 
obtained using the maximum frequency at 
the Kerr ISCO do not change significantly.}. 
The waveform depends on 7 parameters, $({\cal M},\eta,t_c,\phi_c,\chi_s,\chi_a,\delta \varphi)$, 
where ${\cal M}$ and $\eta$ are the chirp mass 
and the symmetric mass ratio, $(t_c,\phi_c)$ 
the time and phase at the coalescence, 
$\chi_{s,a}=(\chi_{1}\pm\chi_{2})/2$ 
combinations of the individual spin components. 
The phase shift enters the the frequency-domain template as $\tilde{h}(f)=Ae^{i\varphi_{\rm GR(f)}}e^{i\psi(f)}$, where 
\begin{align}
\varphi_{\rm GR}=2\pi ft_c&-\phi_c
-\frac{\pi}{4}\nonumber\\
&+\frac{3}{128}(\pi{\cal M}f)^{-5/3}\sum_{i=0}^9\varphi_i(\pi M f)^{i/3}\ ,
\end{align}
is the GR phase (see Appendix A of 
\cite{Pratten:2020fqn} for the explicit 
form of the PN coefficients $\varphi_i$) 
and 
\begin{align}
\psi&=\frac{3}{128\eta^{n/5}}\varphi_{n}\delta\varphi_n
(\pi{\cal M}f)^{(n-5)/3}\quad {\rm if} \quad \varphi_n\neq0\ ,\\
\psi&=\frac{3}{128\eta^{n/5}}\delta\varphi_n
(\pi{\cal M}f)^{(n-5)/3}\quad {\rm if} \quad \varphi_n=0\ .
\end{align}
For the Fisher analysis we 
average over the angles that define the 
source position in the sky, and consider 
Gaussian priors on $\chi_{1,2}$, centered around 
the injected values,  
and with unit width. Finally we 
assume for ET a single L-shaped detector with 
15 km arm-length \cite{Branchesi:2023mws}.

We also show in Fig.~\ref{fig:bound_delta_phi} the constraint on dipole emission, inferred from observations of 
the double pulsar PSR J0737–3039 \cite{Kramer:2021jcw}. 
Such binaries evolve in a low-dynamical regime with 
$v\approx 2\times 10^{-3}$, and provide the tightest 
bound on the -1PN phase deviation $\delta \varphi$.

\section{Accuracy of the trajectory integration}\label{app:ode_error}

Gravitational wave observations constrain the frequency evolution of the waveform with great precision. Therefore, it is key to check that the phase evolution is not affected by systematic errors, which could influence the parameter reconstruction.
We study the accuracy of the numerical integration of the ordinary differential equations (ODEs)~\ref{eq:EMRIODE}. These ODEs are solved using \citep{PRINCE198167} with adaptive step size \url{gsl_odeiv2_step_rk8pd} provided by \cite{gnu_library}.

Firstly, we cross-checked the phase evolution of the implementation against \texttt{Mathematica} on 4 test trajectories.
Then, we investigated the difference of the final phase of an inspiral in GR with an ODE absolute error of $10^{-14}$ with respect to the final phase of an inspiral with a given scalar charge $d$ and various different ODE absolute errors. We show the results in Figure~\ref{fig:phase_difference} for a system with $M = 10^6\,M_\odot$, $\mu = 10\,M_\odot$, $a = 0.95$, $p_0 = 8.343242843079224$, $e_0 = 0.4$ until its plunge.
\begin{figure}
    \centering    \includegraphics[width=\columnwidth]{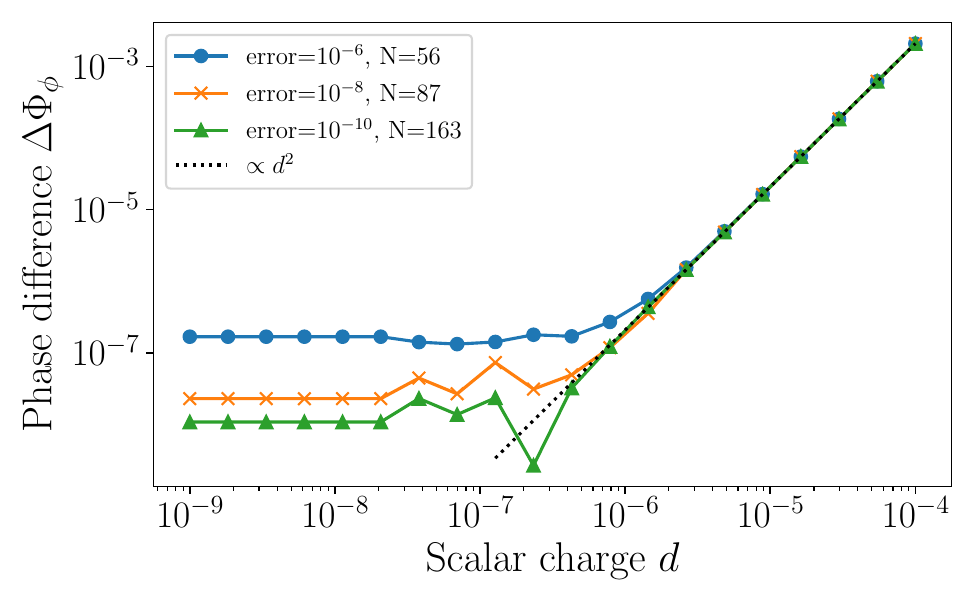}
    \caption{Difference in the final phase 
     between an EMRI evolution obtained with a 
     non-zero scalar charge and various ODE 
     errors (see legend), and an EMRI evolution in GR  with 
     an ODE error of $10^{-11}$.
     For reference, we provide the number of points N taken from the ODE integrator.
     }
    \label{fig:phase_difference}
\end{figure}
For small scalar charges, $d<3\times 10^{-6}$, the phase difference is determined by the ODE solver's noise floor.
On the contrary, the value of the phase difference is independent of the ODE error for large scalar charges $d>3\times 10^{-6}$ and it follows the $d^2$ scaling expected by an expansion of the phase difference for small $d$. This expansion is a good approximation up to $d\approx 1$. The ODE error adopted in this work was $5\times 10^{-10}$.

\section{Systematic bias due to non-zero scalar charge}\label{app:bias}

In this appendix we provide further details on the 
systematic bias that could potentially affect 
EMRI analyses due to the mismatch between a GR 
recovery template, and a signal with a non-zero 
scalar charge. 
This is critical as the charge can influence the 
binary phase evolution and bias parameter estimation. 
To investigate such bias we inject an EMRI signal 
with $d=0.025$ and source parameters $M=10^5\,M_\odot$, 
$\mu=5\,M_\odot$, $a=0.95$, $e_0=0.4$, $T=2$ yrs. 
We analyse the data with two templates: (i) a GR 
waveform, and (ii) a waveform model in which 
the scalar charge is free to vary. The posterior 
distributions on the EMRI parameters are shown in 
Fig.~\ref{fig:bias}. We find 2/3 sigma systematic biases 
in the intrinsic parameters recovered with the GR template. 
{Instead, extrinsic parameters present biases smaller than 1 sigma, and are less affected by the presence of the scalar charge.}
The best log-likelihood point obtained with 
the GR template is $\ln p(s|\Theta)\approx -9$.

\begin{figure*}
    \centering
\includegraphics[width=2\columnwidth]{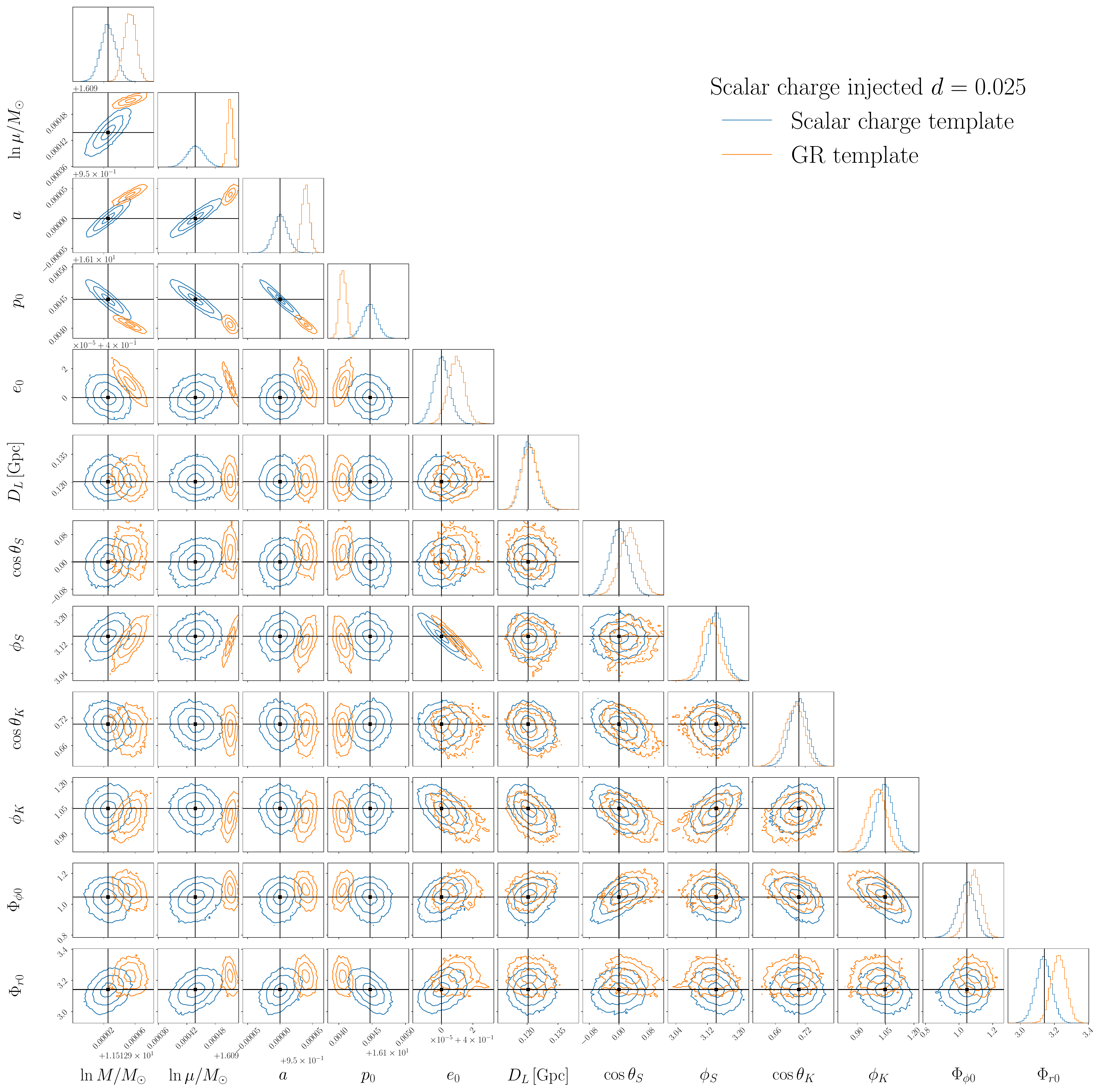}
    \caption{
    Posterior distribution of an EMRI system with scalar charge $d=0.025$ and source parameters $M=10^5\,M_\odot$, $\mu=5\,M_\odot$, $a=0.95$, $e_0=0.4$, $T=2$ yrs and SNR=50. The scalar charge template (blue) correctly recovers the injected parameters, whereas the GR template is biased due to the non-zero scalar charge $d=0.025$. 
    {A downloadable version of this plot can be found \href{https://github.com/lorenzsp/testGRwEMRIs/tree/main/plot_paper/figures}{here}.}
    }
    \label{fig:bias}
\end{figure*}

\bibliographystyle{utphys}
\bibliography{Ref}

\end{document}